\documentclass[pdflatex,sn-nature,iicol]{sn-jnl}% Nature Portfolio reference style; iicol = two-column layout
\usepackage{graphicx}
\usepackage{amsmath,amssymb}
\usepackage{siunitx}
\usepackage{booktabs}
\usepackage{xcolor}
\usepackage{lineno}
\usepackage{bbding}
\usepackage{fontawesome5}

\graphicspath{{figures/}}
\newcommand{\todo}[1]{\textcolor{red}{[#1]}}
\newcommand{\nuone}{\nu_{1}}
\newcommand{\Tstar}{T_{2}^{*}}
\newcommand{\Techo}{T_{2}^{\mathrm{H}}}
\newcommand{\BIR}{BIR-4}
\newcommand{\WURST}{WURST-20}
\newcommand{\up}{{\uparrow}}
\newcommand{\dn}{{\downarrow}}
\newcommand{\Pup}{P_{\up}}
\providecommand{\ket}[1]{|#1\rangle}

\begin{document}

\title[Calibration-insensitive pulses]{Mitigating drift in spin-qubit control with calibration-insensitive pulses}

\author[1]{\fnm{Mridul} \sur{Pushp}}\equalcont{These authors contributed equally to this work.}
\author[2]{\fnm{Jonathan M.} \sur{Hess}}\equalcont{These authors contributed equally to this work.}
\author[1]{\fnm{Seong Woo} \sur{Oh}}
\author[1]{\fnm{Vatsal K.} \sur{Bandaru}}
\author[3]{\fnm{Shize} \sur{Che}}
\author[1]{\fnm{Noah D.} \sur{Johnson}}
\author*[1,3]{\fnm{Anthony J.} \sur{Sigillito}}\email{asigilli@engineering.upenn.edu}

\affil[1]{\orgdiv{Department of Electrical and Systems Engineering}, \orgname{University of Pennsylvania}, \orgaddress{\city{Philadelphia}, \state{Pennsylvania} \postcode{19104}, \country{USA}}}
\affil[2]{\orgdiv{Department of Physics}, \orgname{University of Pennsylvania}, \orgaddress{\city{Philadelphia}, \state{Pennsylvania} \postcode{19104}, \country{USA}}}
\affil[3]{\orgdiv{Department of Computer and Information Science}, \orgname{University of Pennsylvania}, \orgaddress{\city{Philadelphia}, \state{Pennsylvania} \postcode{19104}, \country{USA}}}

\abstract{Silicon spin qubits offer fault-tolerant control fidelities in a platform that is compatible with commercial foundries, but device calibration drifts and control parameters must be retuned every few hours. Here, we replace conventional square control pulses with two drift-resilient pulse families: the adiabatic \BIR{} and the composite CORPSE. In a Si/SiGe quantum-dot device, we benchmark single-qubit gate sets built from each pulse against deliberate miscalibration. We find that \BIR{} offers a $99\%$ Clifford fidelity over a factor of 2 change in drive power and CORPSE maintains similar fidelity over $\pm0.9$\,MHz of offset. When fully calibrated CORPSE offers a $99.97\%$ fidelity per gate, on par with a calibrated square pulse. Starting with tuned pulses, we then track errors using gate-set tomography for ten days without recalibration. \BIR{} holds its rotation angle to $0.3^\circ$ throughout, including across a \SI{1.3}{MHz} frequency jump.  Our results demonstrate that these pulses can maintain high-fidelity operations without constant recalibration, mitigating the need for constant tuning, which is a key bottleneck in scaling quantum-dot-based quantum processors.}


\maketitle

% =====================================================================

% =====================================================================
Quantum-dot spin qubits offer high control fidelities, fast readout, and compatibility with state of the art electronics foundries making them a leading candidate for a scaled quantum computer\cite{Loss1998,Zwanenburg2013,Burkard2023,Zwerver2022,GonzalezZalba2021Scaling}. In the Loss--DiVincenzo architecture, each spin is rotated by a resonant microwave pulse\cite{Loss1998} whose amplitude and frequency must be precisely controlled. The progress of spin qubits over the past decade, from proof-of-principle demonstrations\cite{Petta2005,Zajac2018,Yoneda2018,Watson2018,Xue2019} to fully tuned processors with fault-tolerant control fidelities\cite{Noiri2022,Xue2022,Philips2022,Mills2022,Tarucha2025}, has rested on frequent recalibration of these two control axes; in every case the reported fidelities are those of a qubit whose resonance and Rabi frequencies were measured immediately before and throughout the benchmarking experiment. Both parameters vary from dot to dot and wander in time with charge noise and temperature\cite{Undseth2023}. This is particularly true for micromagnet-equipped devices that enable fast electric-dipole spin resonance (EDSR), where the field gradient that provides the drive increases their sensitivity to the electron's position.
\begin{figure}[!tb]
  \centering
  \includegraphics[width=\linewidth]{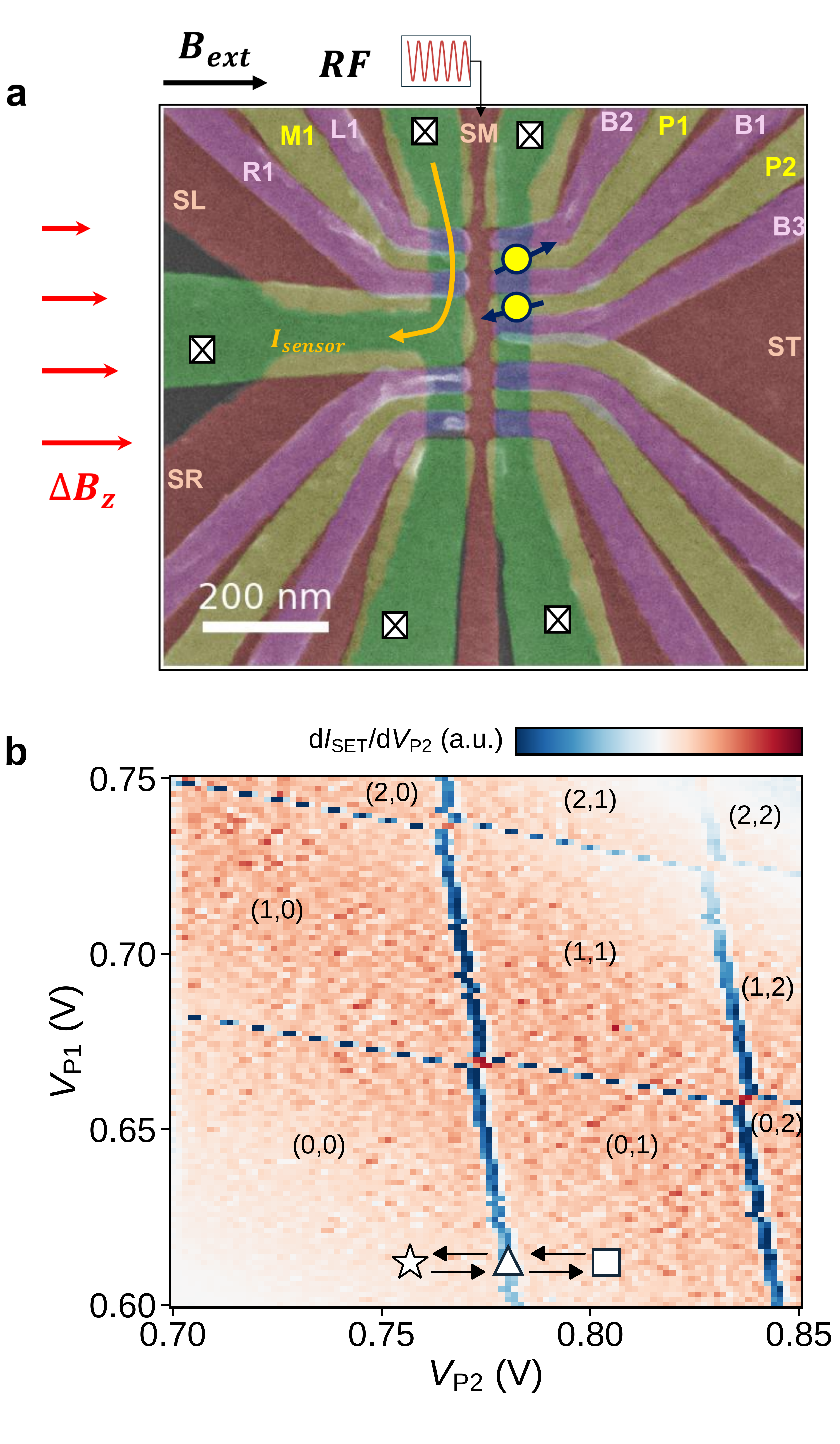}
  \caption{\textbf{Device and readout.}
  \textbf{a}, False-color scanning electron micrograph of a device nominally identical to the one measured. The qubit measured in this work is accumulated under $P2$ and the charge state is measured using a sensor dot accumulated under $M1$.
  \textbf{b}, Charge stability diagram of the double quantum dot, shown as the derivative of the sensor current with respect to $V_{\rm P2}$, with electron occupations $(N_1,N_2)$ labeled. Each experimental sequence starts from the \faStar[regular] point. The qubit is operated with microwaves for spin manipulation in the (0,1) regime at the $\square$ point. Spin selective initialization and Elzerman readout performed at the $\triangle$ point.}
  \label{fig:device}
\end{figure}
Beyond device noise, control instrumentation drifts as well, and in a multi-qubit processor the error sources compound. A multi-qubit controller may deliver a power that depends on how many qubits are addressed, off-resonant drive on neighbors produces Stark shifts\cite{Tarucha2025}, and heating can lead to Zeeman frequency shifts\cite{Undseth2023}. In the laboratory these uncertainties are removed by meticulous and frequent calibration. However, recalibration routines grow with processor size, and this overhead will limit the utility of scaled qubit architectures\cite{Zwolak2026Overcoming}.

The nuclear magnetic resonance (NMR) community faced similar problems decades ago---an inhomogeneous radio-frequency field over a macroscopic sample with a distribution of Larmor frequencies---and solved them with pulses whose rotation angle is insensitive to the field amplitude or to the resonance offset\cite{Levitt1986,Baum1985,Garwood2001,Sigillito2014,Tannus1997,Cummins2000,Cummins2003,Garwood1991}. In spin qubit devices, such pulses have been adopted to improve the fidelity of adiabatic inversion \cite{Laucht2014}. In other work, gradient ascent pulse engineering was used to bolster gate fidelity\cite{Yang2019}. Here, we adopt two families of pulses for spin-qubit control, each designed to mitigate a different calibration axis. \BIR{}\cite{Garwood1991,Tannus1997,Garwood2001} ($B_1$-Insensitive Rotation 4) is an adiabatic pulse, in which the drive frequency is swept through resonance slowly enough for the spin to follow the effective field. Its rotation angle is set by a geometric phase jump and is therefore independent of the pulse amplitude and duration. CORPSE (Compensation for Off-Resonance with a Pulse SEquence)\cite{Cummins2000,Cummins2003} is a composite of three square segments whose middle segment reverses the sign of rotation. The off-resonant phase accumulation from the positive rotation is canceled to first order by the middle, negative rotation. In both cases, drift-insensitivity comes at the cost of longer pulses that are more sensitive to stochastic errors. CORPSE is $\sim4$--8 times longer than its comparable square pulse and \BIR{} is 16--32 times longer.

\begin{figure*}[!tb]
  \centering
  \includegraphics[width=\textwidth]{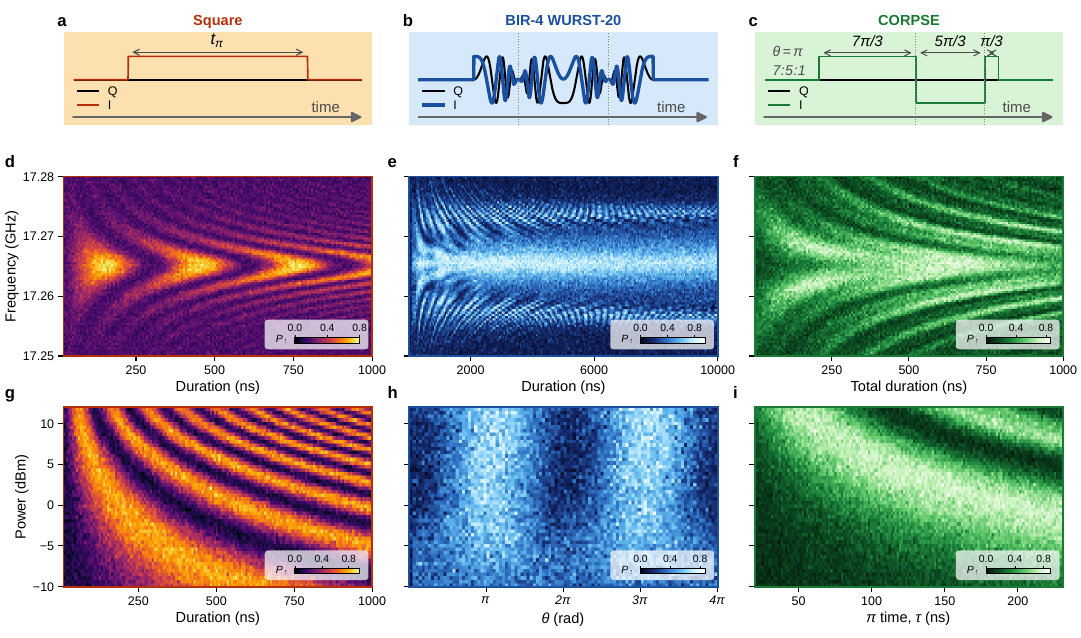}
  \caption{\textbf{Three pulse families and their response to duration, frequency and power.}
  \textbf{a--c}, Schematic I (colored) and Q (black) waveforms of square (\textbf{a}), \BIR{} (\textbf{b}), and CORPSE (\textbf{c}) $\pi$ pulses.
  \textbf{d--f}, Measured spin-up probability $\Pup$ after one pulse versus pulse duration and drive frequency at the nominal power ($\nuone=\SI{3.2}{MHz}$) for the square (\textbf{d}) showing the characteristic chevron pattern. \BIR{} (\textbf{e}) is plotted to a duration of \SI{10}{\micro s}, showing the diabatic regime below $\sim\SI{2}{\micro s}$ and the flat inversion band above it. The side bands appear at the $\pm\SI{10}{MHz}$ sweep limits. CORPSE (\textbf{f}) shows an overlapping chevron pattern characteristic of composite pulses that is broader than the square chevrons shown in (\textbf{d}).
  \textbf{g--i}, $\Pup$ measured while varying drive power and duration for the square pulse (\textbf{g}), tipping angle for the \BIR{} pulse ($\theta$ at fixed $T=\SI{2.5}{\micro s}$, \textbf{h}) and length of a $\pi$ segment for CORPSE (\textbf{i}). The \BIR{} response is independent of power above the adiabatic threshold. Color scales, $\Pup$ as measured.}
  \label{fig:pulses}
\end{figure*}
Here we use these pulse shapes not only to raise the fidelity of a calibrated gate but also to remove the need to maintain calibration. On a silicon quantum-dot spin qubit, we build complete single-qubit gate sets from each pulse, use trains of up to 201 identical pulses (202 $\pi$/2 pulses) to amplify and identify the residual coherent errors, compare the results with simulations, benchmark the gate sets against deliberate frequency and power offsets, and track all three gate sets with gate-set tomography (GST) for ten days. Rather than measuring drift and calibrating it out, these results point to single-qubit control that is robust to drift.

\subsection*{Device and pulse family}
The device is a gate-defined double quantum dot in a Si/SiGe heterostructure with a cobalt micromagnet and a proximal single-electron-transistor (SET) charge sensor~\cite{Sigillito2019} (Fig.~\ref{fig:device}a). A single electron is confined under the $P2$ plunger gate in the (0,1) charge configuration (Fig.~\ref{fig:device}b) and the spin state is measured by spin-selective tunneling to the reservoir~\cite{Elzerman2004}. An in-plane external field of \SI{0.42}{T}, together with the offset field of the micromagnet, gives a Larmor frequency of $f_0=\SI{17.2645}{GHz}$, and EDSR drives spin rotations at a Rabi frequency $\nuone\approx\SI{3.2}{MHz}$ (Supplementary Table~S5). The dephasing time is $\Tstar=\SI{5.8}{\micro s}$ and the Hahn-echo coherence time is $\Techo=\SI{93}{\micro s}$ (Supplementary Note~2), so that the dominant noise is quasi-static on the timescale of a gate. All benchmarking uses Clifford randomized benchmarking\cite{Magesan2011} or gate-set tomography\cite{Nielsen2021}, which are insensitive to state-preparation-and-measurement errors (Methods).

The three waveforms compared throughout are drawn in Fig.~\ref{fig:pulses}a--c. The square pulse, the standard choice for spin qubits~\cite{Pla2012,Veldhorst2014,Kawakami2014}, is a constant-amplitude burst whose rotation angle is proportional to the pulse area, $\theta=2\pi\nuone t$. Amplitude error changes the tipping angle $\theta \propto \sqrt{P}$, and a detuning $\delta$ tilts the rotation axis out of the equatorial plane by $\tan^{-1}(\delta/\nuone)$. Both errors are removed by calibration and are subject to drift.

The \BIR{} pulse (Fig.~\ref{fig:pulses}b) consists of four adiabatic half passages in which the drive frequency is swept linearly under a \WURST{} amplitude envelope\cite{Kupce1995} (Methods). Between passages, where the envelope is zero, the carrier phase jumps by $\pm\Delta\phi$ with $\Delta\phi=\pi+\theta/2$, and the net effect is a rotation by $\theta$ about the initial axis. The tipping angle is thus encoded in the phase jumps rather than in the pulse area, and is a geometric property of the closed loop traced by the effective field~\cite{Berry1984,Jones2000}. The only amplitude-dependent quantity is the adiabaticity $Q=\pi\nuone^2T/2\Omega_0$\cite{Garwood2001}, where $T$ is the pulse duration and $\Omega_0$ the sweep range; the pulse works once $Q\gtrsim2$, and $Q\approx4$--5 in these experiments (Supplementary Note~1). Off resonance, the rotation angle is preserved, but the rotation axis tilts as for a square pulse (error is effectively a $z$ error). This limits \BIR{} overall robustness to frequency errors, even if the tipping angle remains as programmed.

The CORPSE pulse (Fig.~\ref{fig:pulses}c) replaces a rotation about $+x$ with three square segments about $+x$, $-x$ and $+x$, whose angles are chosen so that the first-order detuning error cancels~\cite{Cummins2003} (Methods). The residual detuning error is second order in $\delta/\nuone$, while the amplitude sensitivity is comparable to a square pulse of the same tipping angle. The cost of utilizing CORPSE is the pulse duration. In these experiments, a CORPSE $\pi$ rotation lasts $\approx\SI{680}{ns}$, compared to \SI{157}{ns} for the square pulse. \BIR{} is the longest of the three pulses, taking \SI{2.5}{\micro s} to execute in our experiments. Owing to the waveform granularity of our AWG, the CORPSE gate set uses the $\pi/2$ composite as its only primitive, with each $\pi$ rotation executed as two composites (Methods).

Single-pulse maps show these properties directly (Fig.~\ref{fig:pulses}d--i). Against duration and drive frequency, the square pulse gives the familiar Rabi chevron (Fig.~\ref{fig:pulses}d). \BIR{} responds diabatically, with converging fringes, when the sweep is too short to be adiabatic, and with a flat inversion band above the adiabatic threshold (Fig.~\ref{fig:pulses}e); since the adiabaticity scales as $\nuone^2$, platforms with faster Rabi driving can use correspondingly shorter adiabatic pulses\cite{Garwood2001}. CORPSE gives a chevron broader in frequency than the square pulse's (Fig.~\ref{fig:pulses}f). Against drive power, the square and CORPSE responses fan out into Rabi fringes (Fig.~\ref{fig:pulses}g,i), as expected for pulses without amplitude compensation, whereas for \BIR{} the response to the programmed angle $\theta$, $\Pup=(1-\cos\theta)/2$, is independent of power above the adiabatic threshold (Fig.~\ref{fig:pulses}h).

\begin{figure}[!tb]
  \centering
  \includegraphics[width=\linewidth]{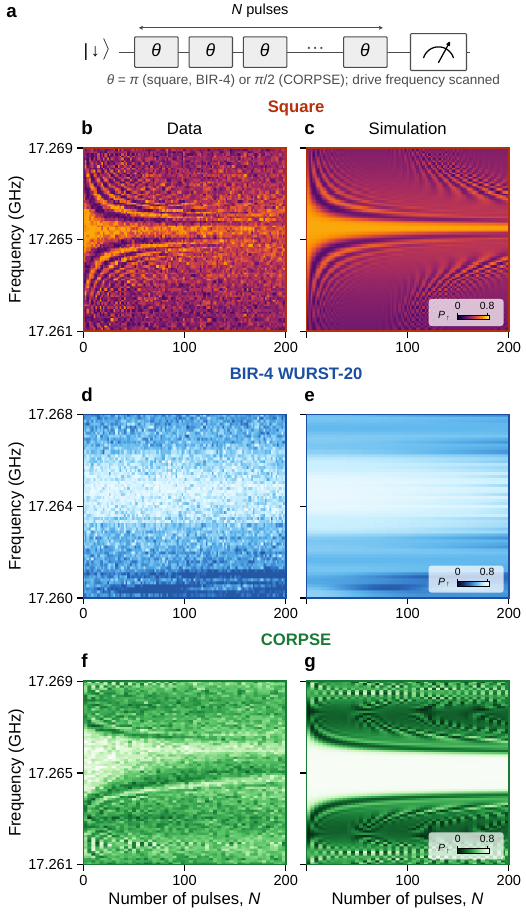}
  \caption{\textbf{Trains of identical pulses versus drive frequency.}
  \textbf{a}, In the pulse sequence, the spin is initialized in $\ket{\dn}$ and $N$ identical pulses ($\pi$ for \textbf{b-e} and $\pi/2$ for \textbf{f,g}) are applied back-to-back as the drive frequency is varied about $f_0$ and $\Pup$ is measured. $N$ chosen to ideally always return spin up. Data is compared against simulation for square pulses(\textbf{b,c}), \BIR{} pulses(\textbf{d,e}), and CORPSE pulses (\textbf{f,g}). Simulations assume a Gaussian quasi-static noise model and assume no free parameters except for readout visibility (Supplementary Note~1). Note that CORPSE and \BIR{} are insensitive to frequency offset compared to conventional square pulses}
  \label{fig:npifreq}
\end{figure}

\subsection*{Amplifying coherent errors with pulse trains}

\begin{figure}[!tb]
  \centering
  \includegraphics[width=\linewidth]{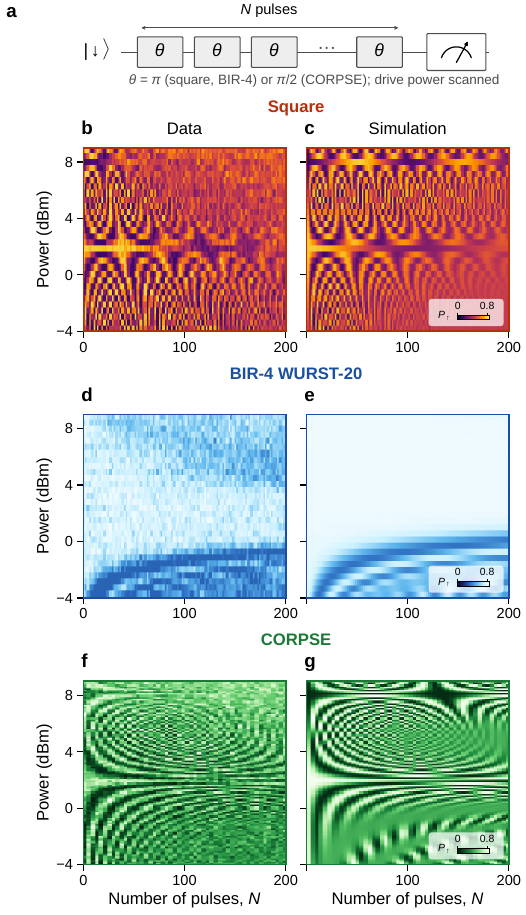}
  \caption{\textbf{Trains of identical pulses versus drive power.}
  \textbf{a}, As in Fig.~\ref{fig:npifreq}
  \textbf{a}, the pulse sequence initializes the spin in the $\ket{\dn}$ state and $N$ identical pulses ($\pi$ for \textbf{b-e} and $\pi/2$ for \textbf{f,g}) are applied back-to-back as the drive power before attenuation is varied and $\Pup$ is measured. $N$ is chosen so the ideal outcome is always $\up$. Data is compared against simulation for square pulses(\textbf{b,c}), \BIR{} pulses(\textbf{d,e}), and CORPSE pulses (\textbf{f,g}). (see supplementary information for a description of the simulation). Note that square and CORPSE pulses are highly sensitive to power, whereas \BIR{} is able to invert the spin regardless of power once the adiabatic condition is satisfied.}
  \label{fig:npipower}
\end{figure}

A train of $N$ nominally identical pulses with no idle between them amplifies a coherent error $\epsilon$ per pulse to $N\epsilon$\cite{Lucero2010}. We apply trains of up to 201 pulses, with $N$ chosen so that the ideal outcome is always $\up$, sweep either the drive frequency (Fig.~\ref{fig:npifreq}) or the drive power (Fig.~\ref{fig:npipower}), and compare each map with a simulation that propagates the Bloch vector exactly through the programmed waveform, averaged over the quasi-static detuning noise set by the measured $\Tstar$, with no fitted parameters apart from the visibility (Methods and Supplementary Note~1).
Under a frequency sweep, the bright band of the square train narrows steadily with $N$, in quantitative agreement with simulation (Fig.~\ref{fig:npifreq}b,c). The \BIR{} band does not narrow (Fig.~\ref{fig:npifreq}d,e): off resonance a \BIR{} pulse still rotates by exactly $\pi$, but does so about a tilted axis. This error amplification experiment, which uses rotations that are multiples of $\pi$, does not accumulate this type of error. The CORPSE band is also largely independent of $N$ (Fig.~\ref{fig:npifreq}f,g), because the in-gate echo cancels the first-order detuning error; its slight asymmetry about resonance reflects a weak frequency dependence of the delivered drive amplitude (Methods and Supplementary Note~4). Under a power sweep the error sensitivity reverses (Fig.~\ref{fig:npipower}): the square and CORPSE trains are bright only along the calibrated-power line and fan into fringes on either side, whereas the \BIR{} train remains bright at every power above the adiabatic threshold. There is some darkening on the \BIR{} train above \SI{4}{dBm} at $N>100$ that was not predicted in the simulation. This is likely due to heating or baseband drive effects.
\subsection*{A calibration map for each pulse}

\begin{figure*}[!tb]
  \centering
  \includegraphics[width=\textwidth]{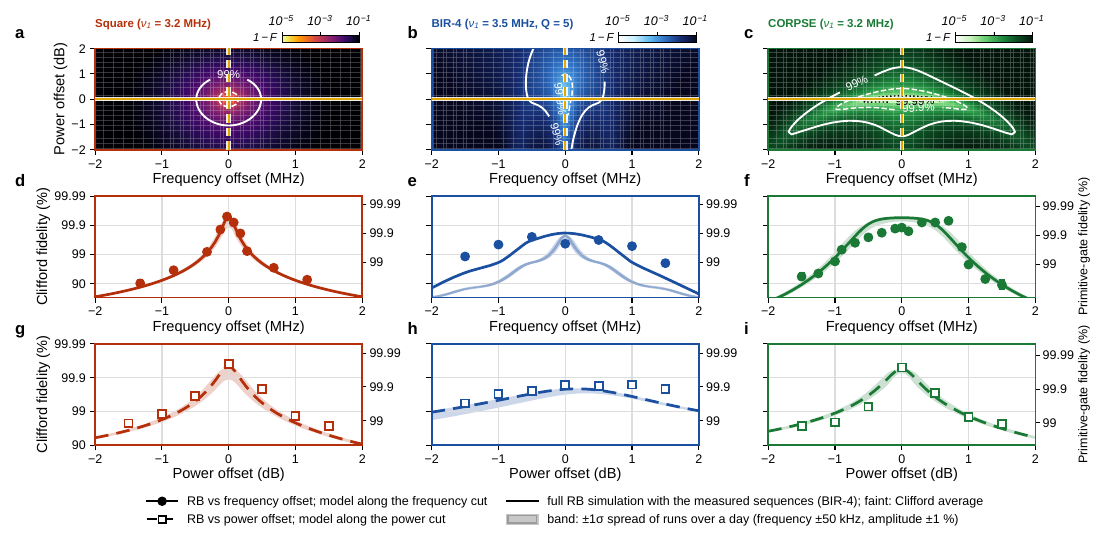}
  \caption{\textbf{Simulated gate infidelity versus frequency and power offset, and the measured RB fidelity along the two offset axes.}
  \textbf{a--c}, Average infidelity per primitive gate, Clifford-weighted over the $\pi/2$ and $\pi$ primitives of the square (\textbf{a}) and \BIR{} (\textbf{b}) gate sets and for the $\pi/2$ composite of the CORPSE gate set (\textbf{c}) plotted as a function of drive-frequency and power offset from the ideal point. The contours mark $99\%$ (solid), $99.9\%$ (dashed) and $99.99\%$ (dotted black) fidelity. The horizontal (solid) and vertical (dashed) lines mark the frequency and power cuts plotted below.
  \textbf{d--f}, Measured Clifford fidelity from randomized benchmarking versus frequency offset (circles); each point is a full RB decay over 100 random sequences at 21 lengths up to 1024 Cliffords. Error bars represent the standard error of the fit. Solid lines represent the simulated fidelity evaluated along the horizontal cut of the map above. For \BIR{} (\textbf{e}) there are two ways to compile rotations with differing sensitivities to detuning error as described in the supplement. The dark curve represents the compilation used in this work (see Supplementary Note~1.10). The \BIR{} on-resonance point is lowered in fidelity compared to off-resonant points due to carrier leakage at baseband (text and Methods).
  \textbf{g--i}, Like panels \textbf{d--f}, except that power is swept. Frequency is fixed near the optimal frequency offset found from panel \textbf{d--f} for each pulse type. The model is shown as a dashed line. Simulations are assembled from measured quantities as outlined in the supplementary information.}
  \label{fig:maps}
\end{figure*}

To turn these observations into a calibration budget we compute, with the same model, the average infidelity of each gate set as a function of frequency and power offset from the calibrated point (Fig.~\ref{fig:maps}a--c, Methods and Supplementary Table~S3) \cite{Nielsen2002}. The three gate families have three different shaped regions of high-fidelity. For the square pulse, the region is a tight circle, requiring both power and frequency to be precisely calibrated. The \BIR{} pulse's region is a vertical band: $99\%$ is maintained over the entire power range above the adiabatic threshold, but its frequency window remains comparable to the square pulse's because the tilted-axis error of a single gate does not cancel as it does in a $\pi$-pulse train. The CORPSE region is a horizontal band, more than twice as wide in frequency as the square window, but with a similar power tolerance. Additionally, it is  the only pulse for which simulations predict a $99.99\%$ region at the experimental drive power, because its in-gate echo suppresses the quasi-static dephasing that limits the other two. Simulations at twice the Rabi frequency (requiring 4 times more power) are found to double every $99\%$ frequency window and lift \BIR{} into the $99.99\%$ regime (Supplementary Table~S3).

\subsection*{Gate fidelity versus calibration error}

We test these predictions with Clifford randomized benchmarking (RB)\cite{Knill2008,Magesan2011} as a function of deliberate frequency and power offsets (Fig.~\ref{fig:maps}d--i; Supplementary Table~S6). The same sequence lengths and fit convention are used for every pulse, and we report the fidelity per Clifford (Methods). The model curves are the Clifford-averaged infidelity of the compiled gate sets along the two cuts through the maps, with no parameters fitted to the RB data other than a single Markovian dephasing floor common to all three pulses (Supplementary Note ~1.4).

\begin{figure*}[!t]
  \centering
  \includegraphics[width=\textwidth]{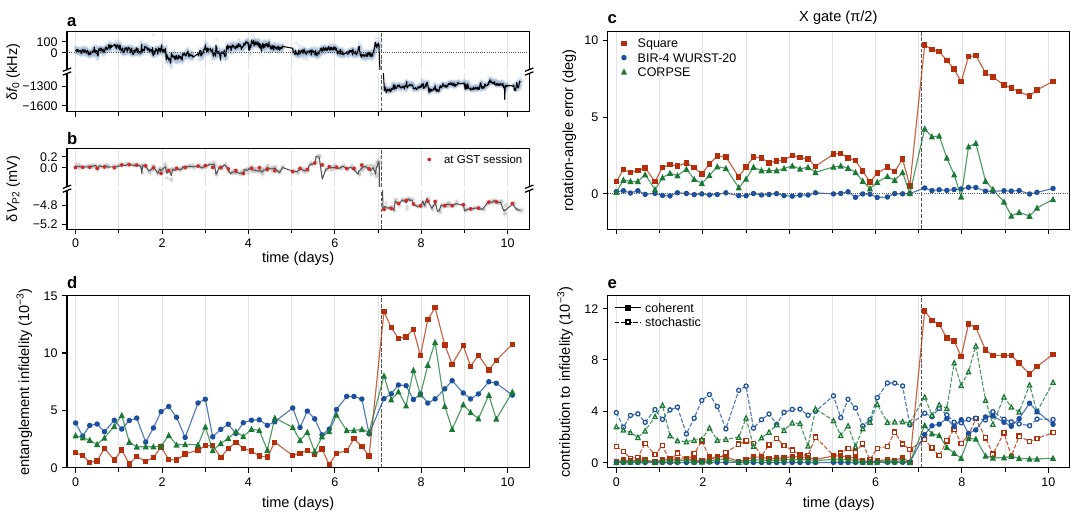}   % = figures/fig6_short.pdf, 183 x 88 mm
  \caption{\textbf{Ten-day GST campaign without recalibration.}
Square, BIR-4 WURST-20 and CORPSE gate sets were characterized by gate-set tomography in 53 sessions per pulse family over ten days (each pulse type was measured every ${\approx}4.5$~h). Each point in \textbf{c}--\textbf{e} represents a full GST measurement, with the vertical dashed line corresponding to a significant charge switching event on day 7. The pulses were calibrated only once at the start of the measurement campaign and only monitored thereafter through Ramsey experiments.
\textbf{a}, Frequency change and \textbf{b}, charge transition voltage change from the initial full calibration on day zero. In each, the dots indicate individual measurements and the dark line is a 10-minute rolling average. Both are plotted with a discontinuous vertical axis because of the charge switching event.
\textbf{c}, Rotation-angle error of the X gate for each pulse type (see Supplementary Fig.~S10 for the Y gate).
The square pulse follows the drift in \textbf{a}, and after the day-7 switch over-rotates by ${\approx}10^\circ$. CORPSE removes all first-order frequency error and keeps rotation errors below $4^\circ$; BIR-4 stays within $0.3^\circ$ throughout.
\textbf{d}, Entanglement infidelity for all gates.
Before the switch, the \SI{2.5}{\micro s} BIR-4 has a constant infidelity of ${\approx}5\times10^{-3}$ while square and CORPSE lie at $(1\text{--}3)\times10^{-3}$. After the switch, square and CORPSE degrade to $(0.6\text{--}1.4)\times10^{-2}$ whereas BIR-4 only rises to ${\approx}8\times10^{-3}$.
\textbf{e}, Decomposition of the X-gate infidelity into its coherent (filled markers, solid lines) and stochastic (open markers, dashed lines) parts as reported by GST.
The coherent error carries the calibration drift and is ordered square $>$ CORPSE $>$ BIR-4 at every session. The stochastic error, a lower bound on the infidelity set by $T_2^*$ and the pulse duration, is largest for the long \BIR{} pulses and smallest for the short square pulses.}
  \label{fig:gst_drift}
  \label{fig:rb}
\end{figure*}
On the frequency axis, CORPSE stays above $99\%$ Clifford fidelity over $\pm0.9$\,MHz of offset, about twice the range of the square pulse, and in both cases the measured window and the roll-off beyond it are reproduced by the model (Fig.~\ref{fig:maps}d,f).

\BIR{} appears more robust still, holding above $99.5\%$ at $\pm\SI{1}{MHz}$ (Fig.~\ref{fig:maps}e), far above the Clifford-averaged model. This is a property of how errors propagate in a randomized benchmarking experiment. Square and CORPSE gates realize negative rotations by a $180^\circ$ shift of the carrier phase, so that the rotation axes are twisted off resonance in four different directions, and the tilt errors of successive gates are added as a random walk. While \BIR{} can use the same approach to realizing negative rotations, here it is encoded as negative differential phase jumps, reducing the effective number of rotation axes to two. When compiling gates using this approach, off resonance errors equate to rotations about a skewed coordinate system (i.e. the $x$ and $y$ axes are not perfectly perpendicular), but gate tipping angles remain exact. A sequence-level simulation reproduces the measured decays (Supplementary Fig.~S4) and we plot the simulated fidelity curves for both the two- and four-axis compilations in Fig.~\ref{fig:maps} (dark blue is the two-axis compilation used in experiment).

When sweeping power (Fig.~\ref{fig:maps}g--i): \BIR{} stays above $99.4\%$ over the full $\pm1.5$\,dB measured, whereas the square and CORPSE gate sets both fall below $98\%$, degrading at the same rate, as their equal amplitude sensitivity requires.

At its optimum, CORPSE reaches a Clifford fidelity of $99.93\%$, on par with the $99.96\%$ of the calibrated square pulse ($99.97\%$ and $99.98\%$ per primitive gate, respectively) and comparable to the highest values reported for quantum dot spin qubits\cite{Yoneda2018,Yang2019,Mills2022,Lawrie2023,Tarucha2025}. \BIR{} reaches $99.84\%$ Clifford fidelity ($99.91\%$ per primitive), limited by dephasing during its \SI{2.5}{\micro s} duration and when on resonance, carrier leakage. Unlike the other two waveforms, we realize \BIR{} pulses through baseband vector modulation (as opposed to single-sideband modulation), so the modulator's leakage is a weak, continuous tone at $f_0$ (Methods and Supplementary Note~4.4). The more detailed error analysis of gate-set tomography (GST) resolves these sources of infidelity\cite{Nielsen2021}.

\subsection*{Ten day continuous GST}
To measure drift tolerance in a realistic setting rather than through deliberate offsets, we characterized all three gate sets by GST in 53 sessions per pulse family over ten days, with the drive frequency, pulse amplitudes, and readout point set once at the start and never adjusted (Fig.~\ref{fig:rb}). During the first seven days the qubit frequency wandered by $\pm\SI{100}{kHz}$ and the drive amplitude by $\pm1\%$. A charge rearrangement then shifted $f_0$ by \SI{-1.3}{MHz} (Fig.~\ref{fig:rb}a,b), and the drive was deliberately left detuned for the remaining three days. The coherent (Hamiltonian) part of each gate's error generator\cite{BlumeKohout2022} carries the calibration drift. After the rearrangement, the square pulse over-rotates by $\approx10^\circ$, the value expected for this detuning (Fig.~\ref{fig:rb}c). Even before the shift, its rotation angle oscillates with a 24-hour period by about $2^\circ$, likely due to drift in the microwave electronics driven by daily temperature fluctuations in the laboratory.

CORPSE tracks the square pulse's amplitude error during the slow drift as expected since both have equal amplitude sensitivity. However, CORPSE is far less affected by the frequency step, retaining a $4^\circ$ error consistent with a detuning just outside its compensated window.

\BIR{} stays within $0.3^\circ$ of its programmed angle throughout, through both the amplitude drift and the frequency step. Its error after the step appears instead as the predicted tilt of its rotation axis: the angle between its $X$ and $Y$ rotation axes, a gauge-invariant quantity, falls from $90^\circ$ to about $80^\circ$, the value expected from the tilted-axis picture and the same as for the square pulse at the same detuning (Supplementary Note~5). The stochastic error, by contrast, is largest for \BIR{} and smallest for the square pulse (Fig.~\ref{fig:rb}e), growing as the pulse durations increase. These values are upper bounds on the per-gate stochastic error, because GST assigns slowly varying error that accumulates coherently across a circuit to the stochastic channel. The \BIR{} gate set is also the only one whose model-violation statistic grows between short and long germ sequences, indicating some drift in control parameters through a circuit, likely due to heating; this is unsurprising given that \BIR{} sequences are 16-32 times longer than square. The remaining CORPSE error is predominantly stochastic and consistent with the $\Tstar$-set floor rather than with calibration.

% =====================================================================
\section*{Conclusions}

Engineered quantum control pulses can help to mitigate the need for calibration and re-tuning in scaled quantum processors. In this work, we explored two variations of drift-resilient pulses. \BIR{} holds its rotation angle through amplitude drift, overcoming noise from the instrumentation and confinement shifts. With faster gate operations in similar spin qubit platforms\cite{wang2021optimal}, BIR-4 can overcome its limitations of slower drive and should offer higher fidelities (Supplementary Note ~1.6). CORPSE, on the other hand, mitigates frequency drift, which has been observed in similar silicon-based QPUs where heating\cite{Undseth2023,Takeda2018} or changing nuclear spin environments\cite{Yoneda2018} can lead to time-dependent drifts in the qubit frequency. Neither pulse overcomes all sources of drift or error, so one should choose a pulse that mitigates the dominant (or perhaps most costly to calibrate) source of calibration drift. A processor with shared microwave delivery, in which $\nuone$ varies from dot to dot through the lever arm and the local field gradient\cite{Vahapoglu2021,Hansen2021}, is served by the adiabatic pulse; a processor with individually addressed dots whose resonances wander with charge noise and heating\cite{Undseth2023} is served by composite pulses. Both appear to be good candidates to offset errors during simultaneous qubit operations.

Over ten days without recalibration, the dominant source of miscalibration in our system was amplitude drift with a period of $\sim24$ hours, pointing to temperature fluctuations that affect the microwave power delivered to the cryostat. A large charge shift, which has been observed in our system before though it is not common, then presented an opportunity to measure the gates under realistic gross miscalibration. The fidelity of the square pulse dropped by $1\%$, that of CORPSE by about a third as much, and that of \BIR{} almost negligibly, although \BIR{} is limited to a significantly higher stochastic infidelity by its length. In terms of absolute fidelity, both \BIR{} and CORPSE pulses outperformed square pulses after the charge shift.

More exotic pulse families exist and will be the subject of future work. For example, a CORPSE sequence can be nested in a BB1-type amplitude-correcting sequence\cite{Wimperis1994,Brown2004,PhysRevLett.95.200501} to generate a pulse sequence robust to both frequency and amplitude drift. We also note that with only a modest increase in Rabi frequency (factor of 2), \BIR{} pulses are expected to offer a significantly wider operation window and a coherent component exceeding four nines (Supplementary Table~S3). For all of these primitive pulses, higher fidelity operation amounts to improving gate speed (or alternatively improving coherence).  Nevertheless, we make the case that as quantum processors scale beyond a few qubits (where calibration can be meticulously maintained), to deployed, utility scale systems, a better choice of gate set will not be the primitives that deliver the highest possible fidelity, but rather the primitives that offer ``\emph{good enough}" fidelity regardless of drift or miscalibration. This study takes the first steps towards identifying those key primitives and lays the groundwork for calibration-free QPU operation.

% =====================================================================
\section*{Methods}

\subsection*{Device and measurement}
The device is fabricated on a Si/SiGe heterostructure with a 7-nm $^{28}$Si quantum well (residual $^{29}$Si 800ppm) grown by chemical vapor deposition at HRL Laboratories, L.L.C. The device uses three overlapping layers of Al gates defining a four-dot linear array and a single quantum dot charge sensor (Fig.~\ref{fig:device}a). The device was tuned to a double quantum dot regime for the work presented here. Measurements were performed in a dilution refrigerator at a base temperature of $\sim\SI{7}{mK}$ (electron temperature \SI{112\pm16}{mK} from the width of the charge transition with a lever arm of \SI{0.15}{eV/V}, Supplementary Note~3) with an in-plane external magnetic field of \SI{0.42}{T}. Spin initialization and readout use energy-selective tunneling \cite{Elzerman2004} with a readout window of \SI{13}{ms} for randomized benchmarking and calibration, decreased to \SI{4}{ms} for GST. The readout visibility, extracted from the extrema of Rabi oscillations, gives $P_{\rm meas}=b+a\Pup$ with $a=0.63$, $b=0.11$ for the calibration data. When running RB experiments, individual $a$ and $b$ values are extracted from Ramsey measurements taken in parallel with RB (Supplementary Note~1). $\Tstar=\SI{5.8}{\micro s}$ and $\Techo=\SI{93}{\micro s}$ were measured by Ramsey and Hahn-echo sequences with square pulses (Supplementary Note~2).

\subsection*{Microwave control and waveform generation}
Baseband I and Q waveforms are generated by a Zurich Instruments HDAWG at \SI{2.4}{GSa/s} (\SI{0.417}{ns} per sample, 16-sample waveform granularity, \SI{300}{MHz} sequencer clock) and applied to the I/Q inputs of a vector signal generator (Rohde \& Schwarz SGS100A equipped with SGU100A) whose carrier is set to $f_0$; the I/Q modulation bandwidth is \SI{80}{MHz}. For the square and CORPSE pulses the HDAWG's digital oscillator supplies an intermediate frequency and the carrier is set at that offset from $f_0$. The intermediate frequency (IF) is \SI{46.3}{MHz}. This frequency is chosen to use an integer number of IF cycles for each segment of the CORPSE waveform, averaging out the effect of I/Q imbalance. Some square-pulse measurements, which do not require balancing of three composite segments, used \SI{25}{MHz}.

The sign reversal of the I and Q envelopes between CORPSE segments is an exact $180^\circ$ phase inversion performed digitally. With this scheme, d.c.\ offsets of the AWG outputs land at the carrier, \SIrange{25}{46}{MHz} from the spin, and carrier leakage and image rejection of even uncalibrated mixers ($\sim$30dB) are inconsequential. The \BIR{} waveform is applied at baseband owing to programming constraints in the AWG. Unfortunately, when driving in baseband, the carrier lies exactly at the qubit frequency and leakage leads to errors that accumulate not only over a single pulse, but over the whole experiment (milliseconds). The I and Q channels were calibrated to minimize carrier leakage and gain imbalance, but the residual leakage is visible in the RB data of Fig.~\ref{fig:maps}e, where the on-resonance \BIR{} point lies $0.1$--$0.2\%$ below its $\pm\SI{0.5}{MHz}$ neighbors and varies from day to day with the d.c.-offset calibration of the modulator; the excess of $(1$--$2)\times10^{-3}$ per Clifford corresponds to a leakage Rabi rate of $3$--$\SI{4}{kHz}$, about $-60$\,dB relative to the drive, a normal carrier-suppression level (Supplementary Note~4.4). A PIN-diode switch gates the microwave line after amplification and is left open for the duration of each triggered pulse sequence. Drive power is set by the microwave source or by the amplitude of the IQ waveforms set by the AWG. 

Two properties of the drive chain affect the fidelity of the gate. First, above $\sim\SI{8}{dBm}$ the output amplifier compresses: the measured $\nuone(P)$ follows a smooth saturation of the form $\nuone=\nuone^{\rm lin}/[1+(\nuone^{\rm lin}/\nu_{\rm sat})^{2p}]^{1/2p}$, where $\nuone^{\rm lin}\propto10^{P/20}$ is the uncompressed value, with $\nu_{\rm sat}=\SI{10.8}{MHz}$ (reached at an input of \SI{12.6}{dBm}) and $p=2.4$ \cite{Rapp1991}. This saturation is included in the simulations of Fig.~\ref{fig:pulses}g and of the power cuts of Fig.~\ref{fig:maps}g--i. Second, the power delivered to the device depends on frequency. Reflections at the connectors, bias tees and bond wires of the microwave lines and the sample board form standing waves so that across the $\pm\SI{4}{MHz}$ of our frequency scans the drive amplitude varies approximately linearly, by $2.5\pm0.5$\% per MHz (\SI{0.2}{dB} per MHz), inferred from the asymmetry of the CORPSE train of Fig.~\ref{fig:npifreq}f (Supplementary Note~4.3). This gradient is included in the simulations where stated; it converts part of any frequency drift into a power drift and means that there is some interdependence between the two calibration axes.

Unless otherwise stated, square pulses are tuned to have a duration $t_\pi=\SI{157}{ns}$ ($\nuone=\SI{3.2}{MHz}$). \BIR{} pulses have total duration $T=\SI{2.5067}{\micro s}$ (6{,}016 samples), a \WURST{} envelope $A(t)=1-|\sin(2\pi t/T)|^{20}$, three linear frequency sweeps $0\!\to\!+\Omega_0$, $-\Omega_0\!\to\!+\Omega_0$, $-\Omega_0\!\to\!0$ with $\Omega_0/2\pi=\SI{10}{MHz}$ implemented as quadratic phase ramps in the baseband samples, and phase jumps $\pm\Delta\phi$, $\Delta\phi=\pi+\theta/2$, at the two envelope zeros. The adiabaticity is calculated as $Q=\pi\nuone^2T/2\Omega_0$. A CORPSE rotation by $\theta$ is composed of three square segments with angles $\theta_1 = 2n_1\pi + \theta/2 - k$, $\theta_2 = 2n_2\pi - 2k$, and $\theta_3 = 2n_3\pi + \theta/2 - k$ about $+x$, $-x$ and $+x$, respectively, where $k = \arcsin\!\left[\sin(\theta/2)/2\right]$ and $n_{1,2,3}$ are integers chosen to enforce positive pulse angles~\cite{Cummins2003}. Every member of this family of solutions cancels the first-order detuning error; the second-order term depends partly on $n=n_1-n_2+n_3$. That can be minimized by selecting $n_1=n_2=1$, $n_3=0$, which is the shortest physical combination with $n=0$. For $\theta=\pi$ the segment angles are $7\pi/3$, $5\pi/3$ and $\pi/3$; for $\theta=\pi/2$ they are not rational multiples of $\pi$, $\theta_1\approx384.3^\circ$, $\theta_2\approx318.6^\circ$ and $\theta_3\approx24.3^\circ$. The three segments are written as one waveform with integer sample counts in the exact segment ratios, the 16-sample granularity of the AWG is applied to the whole waveform, and the drive amplitude, not the duration, is the calibrated parameter for fine tuning. The $\pi$ and $\pi/2$ composites cannot both be realized with exact segment ratios at a common amplitude and granularity-compatible length, so the CORPSE gate set uses the $\pi/2$ composite as its only primitive, with $\pi$ rotations executed as two $\pi/2$ composites; the square and \BIR{} gate sets use both $\pi/2$ and $\pi$ primitives (Supplementary Note~1.2). The Rabi frequency on the \BIR{} (baseband) path was measured separately with a square-pulse chevron played through the same path, giving $\nuone=\SI{3.5}{MHz}$ at the nominal amplitude.

\subsection*{Calibration with pulse trains}
The resonance frequency is set by Ramsey interferometry with a \SI{1}{MHz} offset from the expected Larmor frequency. Pulse durations are calibrated with odd-$N$ pulse trains played back-to-back with no idle. For the square pulse, both the $\pi$ and the $\pi/2$ lengths are calibrated this way (the $\pi/2$ train with $N=2 + 4m$, for integer $m$). CORPSE is calibrated only using the $\pi / 2$ variant. After the optimal duration is found, the power and frequency is scanned with the same train to complete the calibration (Fig.~\ref{fig:npifreq} and Fig.~\ref{fig:npipower}).

\subsection*{Randomized benchmarking}
For the square and \BIR{} pulses the 24 single-qubit Cliffords are compiled into the primitive set $\{\pm X_{\pi/2},\pm Y_{\pi/2},X_\pi,Y_\pi\}$, \num{1.875} primitives per Clifford on average \cite{Epstein2014}; for CORPSE they are compiled into $\{\pm X_{\pi/2},\pm Y_{\pi/2}\}$ only, with every $\pi$ rotation executed as two $\pi/2$ composites, \num{2.17} primitives per Clifford on average. $Y$ rotations are implemented by a $90^\circ$ phase offset of the I/Q waveform in all cases. Negative rotations are realized for the square and CORPSE pulses by advancing the carrier phase by $180^\circ$, so that a detuning tilts the $+x$ and $-x$ (or $+y$ and $-y$) rotation axes towards $+z$ in opposite directions. For \BIR{}, whose angle is set by the phase jumps, negative rotation is obtained by reversing the sign of the jump offset ($\Delta\phi=\pi-\theta/2$), which is a rotation by $-\pi/2$ about the \emph{same} tilted axis as the positive rotation. The \BIR{} gate set therefore contains only two physical rotation axes (Supplementary Note~1.9). 

The experimental pulse sequence lists were exported and used directly in the sequence-level simulations. Sequences of 21 lengths, approximately log-spaced from $m=1$ to $1024$ Cliffords, each followed by the recovery Clifford, are averaged over $N_{\rm seq}=100$ random sequences. The survival probability $P(m)$ is fitted to $B+Ap^m$ with $B$ fixed to the mixed states level of a Ramsey reference measured immediately before each run. The fidelity per Clifford is $F_{\rm C}=1-(1-p)/2$, and the fidelity per primitive is $1-(1-F_{\rm C})/n_{\rm p}$ with $n_{\rm p}$ the primitives per Clifford of the respective compilation. The quoted uncertainties are the standard errors of this fit (Supplementary Table~S6). The same RB sequence lengths and fit convention are used for all three pulses and all offsets(Supplementary Note~6). For \BIR{} the simulated coherent floor is $1.7\times10^{-4}$ per gate at $Q\approx5$, to which the Markovian floor $T_{\rm M}=\SI{1}{ms}$ (Numerical simulations) adds $1.6\times10^{-3}$ per Clifford; the measured on-resonance value is therefore dominated by the latter.  The asymmetry of the CORPSE map between positive and negative power offsets arises from the interplay of the amplitude error with the second-order detuning residual: detuning raises the effective Rabi frequency to $\sqrt{\nuone^2+\delta^2}$ and partly cures an under-rotation.

\subsection*{Gate-set tomography}
GST was performed on the gate set $\{I,X_{\pi/2},Y_{\pi/2}\}$ with pyGSTi\cite{Nielsen2021,Nielsen2020pygsti} using the standard fiducials and germs for this set and maximum germ-sequence lengths $L\le64$, 120 shots per circuit, in 53 sessions per pulse family (159 in total). Each pulse family was remeasured every $\sim\SI{4.5}{h}$ over 10 days with measurements over the three pulse families interleaved. The pulses were calibrated at the start of the experiment and left fixed thereafter. Estimates are CPTP-constrained (CPTPLND) and the error generator of each estimated gate, $\mathcal{L}=\log(G\,G_{\rm ideal}^{-1})$ in the Pauli-transfer-matrix basis, is decomposed into its Hamiltonian (antisymmetric) and stochastic (symmetric) parts\cite{BlumeKohout2022} using the PyGSTi package. Fig.~\ref{fig:rb}c--e plot the rotation-angle error of the $X$ gate, the entanglement infidelity of all gates, and the coherent and stochastic contributions to the $X$-gate infidelity. The stochastic part corresponds to an average entanglement infidelity of $4.0\times10^{-3}$ (\BIR{}), $2.6\times10^{-3}$ (CORPSE) and $1.2\times10^{-3}$ (square), with a per-session uncertainty of $\pm1\times10^{-3}$. After the frequency step the angle between the $X$ and $Y$ rotation axes of \BIR{} falls from $90.0$--$90.5^\circ$ to $79.7$--$81.1^\circ$, as expected when both axes tilt towards $z$ (Supplementary Note~5).

\subsection*{Numerical simulations}
All simulations propagate the Bloch vector exactly through the programmed waveform as a sequence of piecewise-constant rotations (a \SI{2}{ns} step size for shaped pulses and one AWG sample step size for square segments). Quasi-static dephasing is modeled as a Gaussian distribution of detunings integrated by Gauss--Hermite quadrature. Two widths appear in this work: an earlier exponential fit of the Ramsey decay gave a width of \SI{70}{kHz}, which was used for the pulse-train maps of Figs.~\ref{fig:npifreq}--\ref{fig:npipower}. Later, a more refined Gaussian fit gave $\sigma=1/(\sqrt2\,\pi\Tstar)=\SI{39.5}{kHz}$ for $\Tstar=\SI{5.7}{\micro s}$, which is used for the infidelity maps and model curves of Fig.~\ref{fig:maps} (Supplementary Note~1). Each primitive is averaged over the ensemble before the Cliffords are composed, and sequence-level simulations draw the detuning once per sequence. The maps of Figs.~\ref{fig:pulses}--\ref{fig:npipower} are simulated on the same frequency, power and pulse-number grids as the measurements with the measured $\nuone$, $f_0$, $\Tstar$ and readout constants and use no fitted parameters. The infidelity maps of Fig.~\ref{fig:maps}a--c evaluate the average gate fidelity $\bar F=[3+\mathrm{Tr}(R_{\rm ideal}^{\rm T}R)]/6$ of the ensemble-averaged SO(3) matrix $R$. The model curves of Fig.~\ref{fig:maps}d--i are the Clifford-averaged infidelity of the 24 compiled Cliffords built from the same primitives, plus a Markovian dephasing floor $(1-e^{-t/T_{\rm M}})/3$ per primitive of duration $t$ with $T_{\rm M}=\SI{1}{ms}$. The maps are per primitive gate (Clifford-weighted over the $\pi/2$ and $\pi$ primitives for the square and \BIR{} pulses; the $\pi/2$ composite alone for CORPSE); per Clifford the infidelities are larger by the number of primitives per Clifford and the $99\%$ windows correspondingly narrower. For \BIR{} the frequency cut is also simulated at sequence level, by propagating the exported waveform sample by sample through the exported RB sequences and fitting the result as the data are fitted (Supplementary Note~1.10). Full details, including waveform sample counts, the noise ensemble, and the hardware-imperfection models, are given in Supplementary Note~1. The simulation code is available as described below.

\subsection*{Data availability}
The data that support the findings of this study are available from the corresponding author upon reasonable request. The data are part of an active research program on this device, and they are not deposited in a public repository in order to avoid their misappropriation, including their scraping for the training of artificial-intelligence models, before the program is complete.

\subsection*{Code availability}
The code used to simulate the spin dynamics will accompany the published manuscript in a public Github repository.

% =====================================================================

% =====================================================================
\clearpage   % flush remaining floats before the back matter
\backmatter

\bmhead{Acknowledgments}
We thank Bryan Thomas, Clayton Jackson, and Joe Kerckhoff at HRL Laboratories L.L.C. for providing the $^{28}$Si heterostructures that made this work possible. This work was funded by the AFOSR YIP program under grant number FA9550-25-1-0143. The authors also acknowledge indirect financial support from the AFOSR under grants FA9550-26-1-B285-DEF and PAM-P25-003241-SA01 as well as the ARO under grant W911NF-23-1-0242. V.K.B acknowledges an NSF GRFP (DGE-2236662).

\bmhead{Author contributions}
M.P., J.M.H., and A.J.S. conceived the idea for the experiments. N.D.J. and S.W.O. fabricated the device. J.M.H. and M.P. performed the measurements with help from S.W.O.,N.D.J.,V.K.B. and S.C. A.J.S. and V.K.B. assisted in simulations. S.C. wrote the qubit control software. M.P., J.M.H. and A.J.S. analyzed the data and wrote the manuscript with input from all authors. A.J.S. supervised the project.

\bmhead{Competing interests}
The authors declare no competing interests.

\bmhead{Additional information}
\textbf{Supplementary information} is available for this paper at \todo{link}.\\
\textbf{Correspondence} and requests for materials should be addressed to asigilli@engineering.upenn.edu

\bibliography{refs}

\end{document}